\documentclass[twocolumn, superscriptaddress, amsmath, amssymb, aps, prapplied, floatfix, longbibliography]{revtex4-2}

\usepackage{graphicx}
\usepackage{dcolumn}
\usepackage{bm}
\usepackage{siunitx}
\usepackage{textcomp}
\usepackage{xcolor}

\begin{document}

\title{A fully packaged multi-channel cryogenic module for optical quantum memories}

\author{David J.~Starling}
 \email{David.Starling@ll.mit.edu}
\author{Katia Shtyrkova}
 \affiliation{Lincoln Laboratory, Massachusetts Institute of Technology, Lexington, MA 02421, USA}
\author{Ian Christen}
 \affiliation{Research Laboratory of Electronics, Massachusetts Institute of Technology, Cambridge MA 02139, USA}
\author{Ryan Murphy}
 \affiliation{Lincoln Laboratory, Massachusetts Institute of Technology, Lexington, MA 02421, USA}
\author{Linsen Li}
\author{Kevin C.~Chen}
 \affiliation{Research Laboratory of Electronics, Massachusetts Institute of Technology, Cambridge MA 02139, USA} 
\author{Dave Kharas}
\author{Xingyu Zhang}
 \thanks{Now at McKinsey \& Company}
\author{John Cummings}
\author{W.~John Nowak}
\author{Eric Bersin}
\author{Robert J.~Niffenegger}
 \thanks{Now at University of Massachusetts, Amherst}
 \affiliation{Lincoln Laboratory, Massachusetts Institute of Technology, Lexington, MA 02421, USA}
\author{Madison Sutula}
\author{Dirk Englund}
 \affiliation{Research Laboratory of Electronics, Massachusetts Institute of Technology, Cambridge MA 02139, USA}
\author{Scott Hamilton}
\author{P.~Benjamin Dixon}
 \affiliation{Lincoln Laboratory, Massachusetts Institute of Technology, Lexington, MA 02421, USA}
 
\date{\today}

\begin{abstract}
Realizing a quantum network will require long-lived quantum memories with optical interfaces incorporated into a scalable architecture. Color centers quantum emitters in diamond have emerged as a promising memory modality due to their optical properties and compatibility with scalable integration. However, developing a scalable color center emitter module requires significant advances in the areas of heterogeneous integration and cryogenically compatible packaging. Here we report on a cryogenically stable and network compatible quantum-emitter module for memory use. This quantum-emitter module is a significant development towards advanced quantum networking applications such as distributed sensing and processing. 
\end{abstract}

\maketitle

\section{\label{sec:intro}Introduction}

Quantum memory systems are a crucial enabling technology for the continued development and realization of advanced quantum network capabilities and applications~\cite{Wehner2018}. Indeed, memory systems scaled to tens or hundreds of memories with millisecond-class coherence times would enable quantum repeater functionality~\cite{Briegel1998,Azuma2022} for improved network performance~\cite{Munro2010,Bhaskar2020,Dhara2022,Lee2022} and quantum network applications such as improved sensing~\cite{Khabiboulline2019} and distributed quantum processing~\cite{Monroe2014,Cuomo2020}.

There are multiple candidate qubit platforms for networked quantum memory use, including trapped ions~\cite{Nichol2022,Krutyanskiy2022} 
and neutral atoms~\cite{Daiss2021,vanLeent2022}, atomic ensembles in vapors~\cite{Luo2022,Wang2022} and in solid-state materials~\cite{Zhong2017,Lago-Rivera2021,Askarani2021}, and individual color centers in solids~\cite{Dibos2018,Kindem2020,Miao2020}. Diamond color-center emitters have emerged as a particularly promising platform due to several beneficial characteristics~\cite{Ruf2021}. Their coherent optical interface provides compatibility with optical fiber networks~\cite{Pompili2021, Stolk2022}. They have the potential to scale to thousands of memory qubits through on-chip integration~\cite{Wan2020}, and their electron spin degree of freedom can exhibit coherence times ranging from milliseconds~\cite{Sukachev2017} to seconds~\cite{Abobeih2018}, providing utility as quantum repeaters and for quantum network applications. Furthermore, ancillary nuclear spins can be used as logical registers~\cite{Bradley2019} for multi-qubit processing such as error detection \cite{Stas2022} and  correction~\cite{Waldherr2014}. However, the challenges of engineering a robust and scalable memory module has limited the use of these memories in network testbeds, delaying demonstration of advanced quantum network applications.

\begin{figure}[t]
	\includegraphics[width=\linewidth]{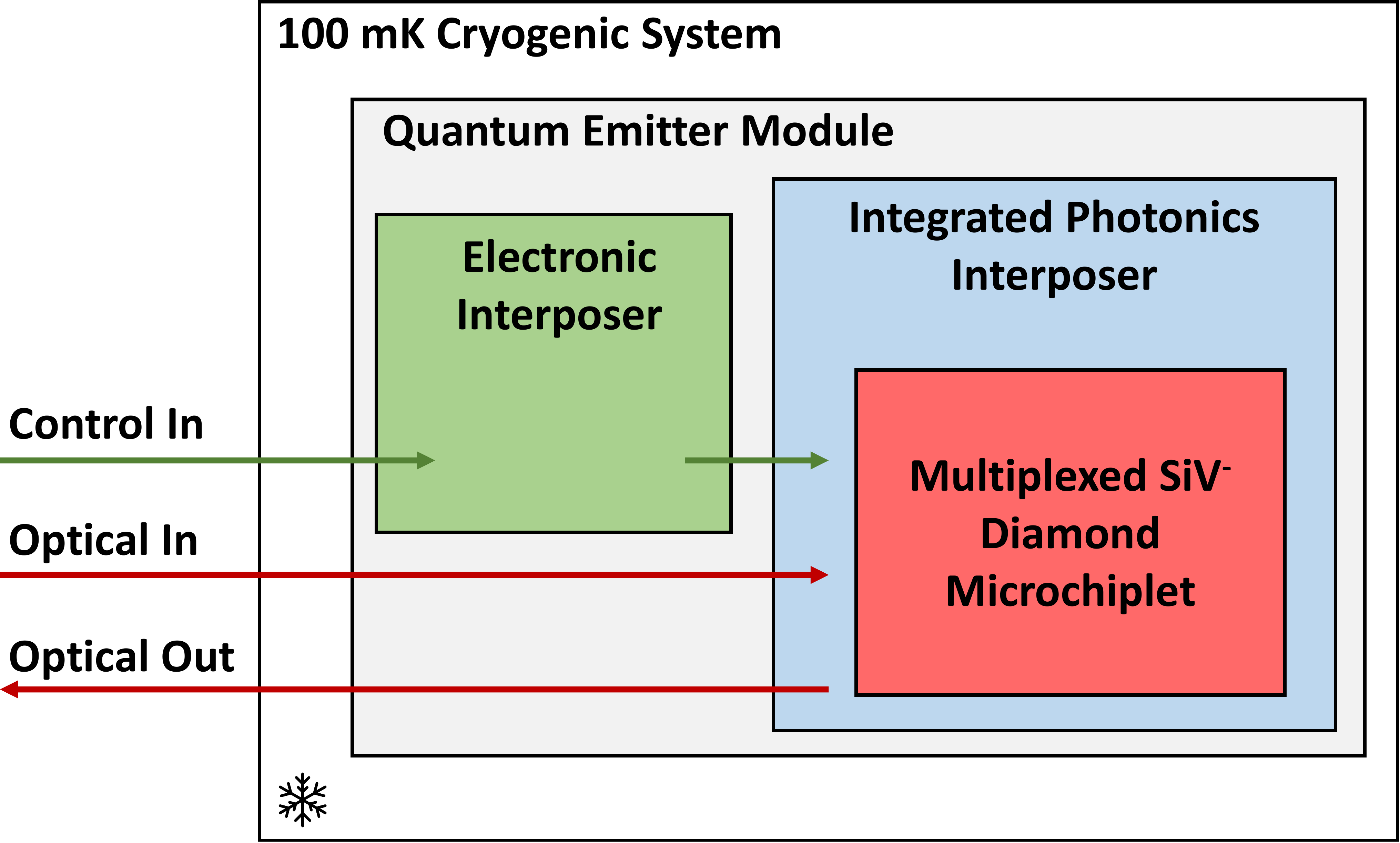}
	\caption{\label{fig:highlevel} A conceptual block diagram of the quantum-emitter module. The module operates in a cryogenic environment. Electrical control signals are fed into the cryogenic system and routed via an electronic interposer that directs signals to an integrated photonics interposer that can interface with a multiplexed SiV${}^-$ diamond microchiplet for qubit control. Optical signals are fed into the cryogenic system and routed via the same photonic interposer to enable optical interaction with the multiplexed SiV${}^-$ diamond microchiplet.}
\end{figure}

Here we report on our efforts to develop a robust, packaged quantum-emitter module. We build upon previous work~\cite{Wan2020,Niffenegger2020,Mehta2020} to demonstrate heterogeneous integration of optical fibers, photonic integrated circuits (PICs) and diamond color centers. Our results rely upon novel custom designed and fabricated silicon-nitride (Si${}_3$N${}_4$, SiN) PICs, matched diamond microchiplets with waveguides containing negatively charged silicon vacancy (SiV${}^-$) color centers, a custom diamond alignment and stamping process, as well as a novel precision fiber alignment and bonding procedure. We follow the previously reported general strategy of multi-waveguide-fiber architecture ~\cite{Wan2020, Niffenegger2020, Mehta2020} and incorporate our developments to realize a multi-emitter module that is fully packaged, cryogenically compatible, inherently scalable, and network-compatible. The resulting quantum-emitter module architecture enables scaling of large numbers of optical quantum memories to be readily integrated into emerging network testbeds~\cite{Cui2021, Du2021, Chung2022} for the realization of applications in quantum networking.

\section{\label{sec:package}Design and Development}

The conceptual goal of our work was to develop a cryo-compatible, fully-packaged module that interfaces optical fiber with a controllable quantum emitter, as shown in Fig.~\ref{fig:highlevel}. To accomplish this task, our design uses a custom electronic interposer, an integrated photonic interposer (the PIC) and a diamond microchiplet---incorporated into a commercial dilution refrigerator. The electronic interposer was produced commercially, and the eight-channel diamond micro chiplet was fabricated using a similar process to the one shown in Refs.~\cite{Mouradian2017, Wan2018, Wan2020} (see Appendix~\ref{app:diamondfab} for details). In what follows, we will focus on the novel fabrication of the integrated photonic interposer and the packaging methodologies. 

\begin{figure}[t]
	\includegraphics[width=\linewidth]{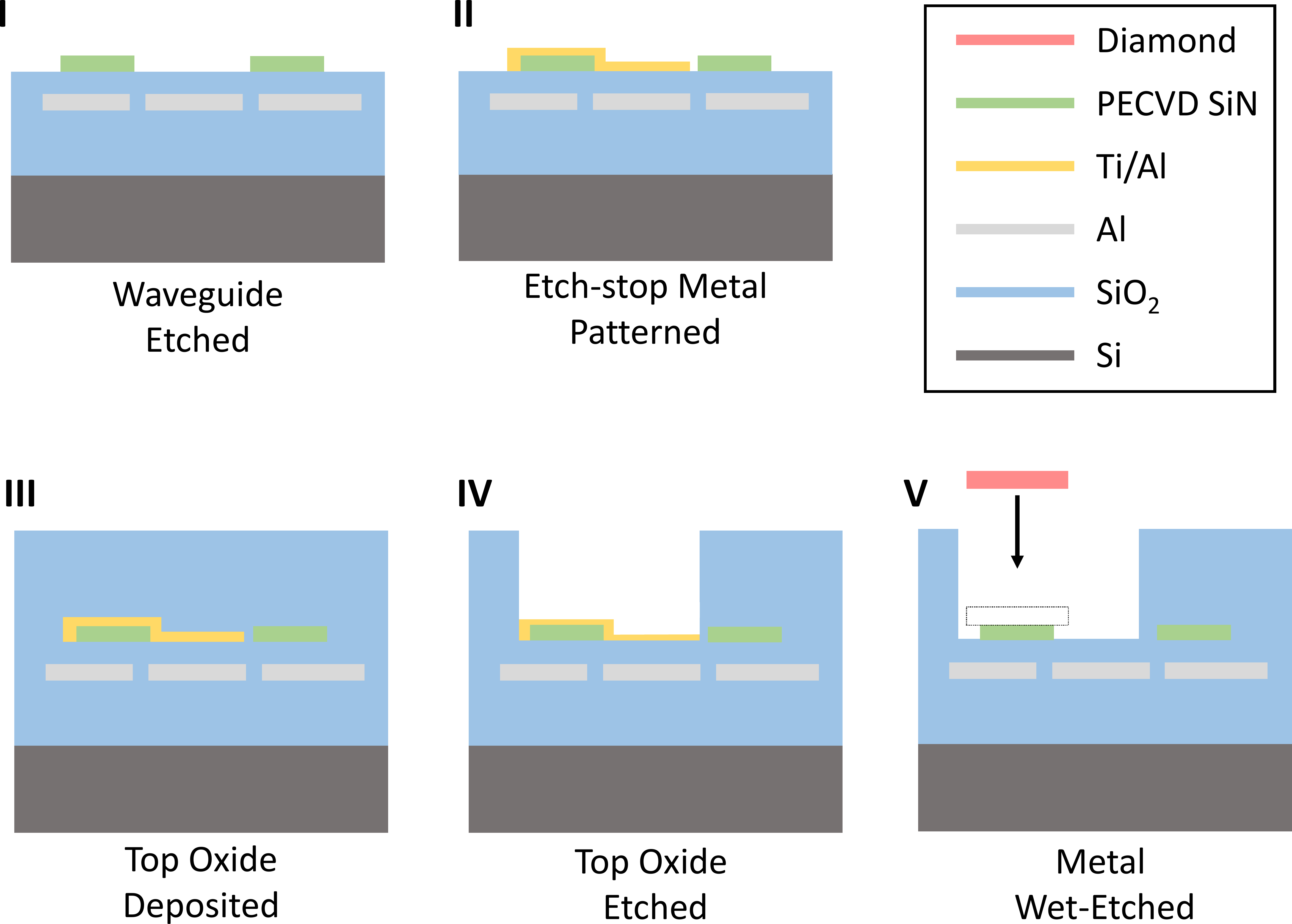}
	\caption{\label{fig:fab} Fabrication process steps relevant to the novel PIC window opening, detailed in the text. Fully clad SiN waveguides allow for efficient mode overlap to standard single mode fiber, while a precision window etch provides for heterogeneous integration of a diamond microchiplet.}
\end{figure}

\begin{figure}[t]
	\includegraphics[width=\linewidth]{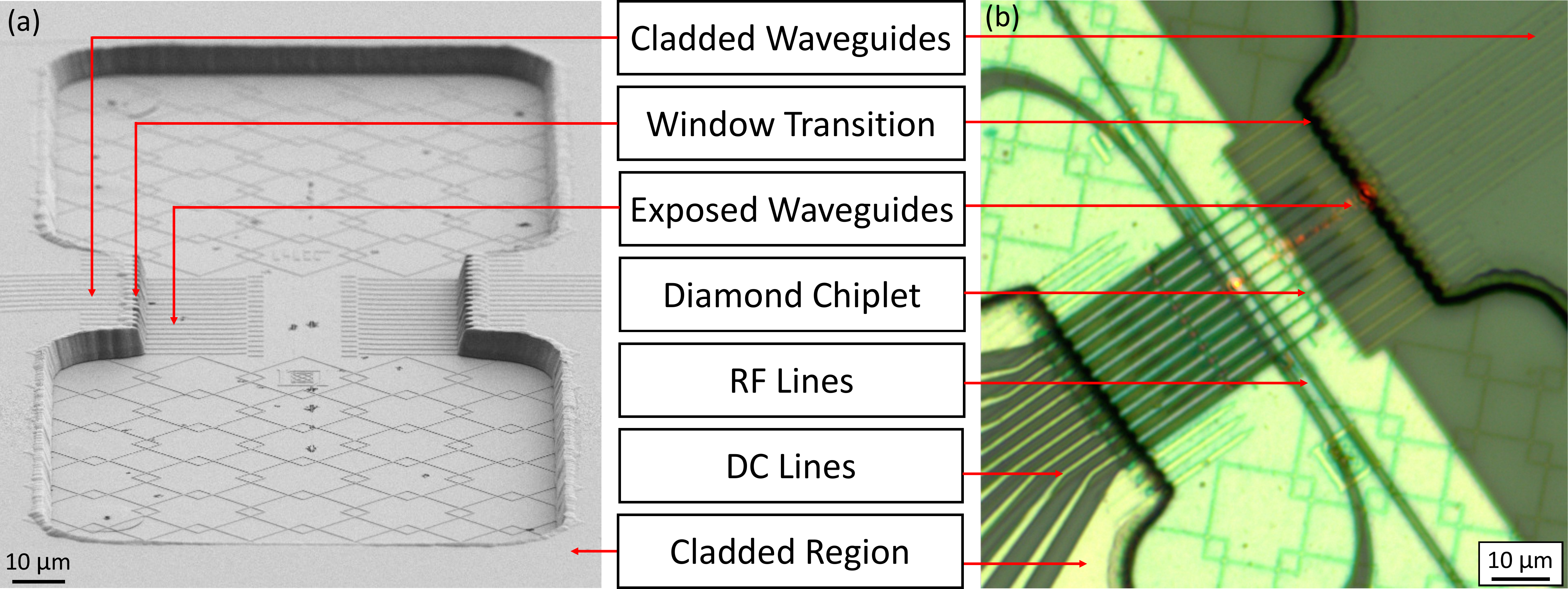}
	\caption{\label{fig:placement} The diamond window. (a) A scanning electron microscope image of the diamond window region with exposed SiN waveguides. (b) An optical image of a stamped diamond micro-chiplet onto the exposed SiN waveguides. Here you can see the alignment of the diamond waveguides as well as the microwave and DC electrical lines which can be used to control the electronic states of the SiV${}^-$.}
\end{figure}

We designed and fabricated a custom PIC interposer to integrate a commercial optical fiber array with a diamond microchiplet. This process has been accomplished previously in Ref.~\cite{Wan2020} using unclad aluminum nitride waveguides on a sapphire wafer where the waveguides were exposed (unclad) to allow for heterogeneous integration. However, the small optical mode in the unclad aluminum nitride waveguide at the facet necessitated the use of lensed and/or high numerical aperture fiber to obtain efficient coupling. Additionally, it is possible to use a thin layer of cladding followed by a precision etch \cite{Chanana2022}, but fabrication tolerances limit the depth of the window. Thick cladding, however, is critical to produce a large and symmetric mode well-coupled to cleaved optical fiber. Here, we demonstrate a silicon nitride (SiN) fabrication process that allows for both waveguides with thick oxide cladding at the facet and waveguides without oxide cladding for diamond integration. As shown in Fig.~\ref{fig:fab}, we employed a titanium-aluminum etch-stop layer directly on top of the SiN waveguides used for diamond integration, allowing a standard oxide etch to open a precision window in the top SiO${}_2$ cladding without damaging the SiN layer (see Appendix~\ref{app:PICfab} for details). The metal was then removed with a selective wet etch, and the resulting open window is shown in Fig.~\ref{fig:placement}(a). The oxide-clad waveguides at the facet allows for a larger optical mode matched to cleaved optical fiber resulting in low-loss coupling to a 20-channel commercial fiber array with 630HP optical fiber.

The PIC was designed to support eight independent quantum-emitter channels, based upon the eight independent diamond waveguides available on a single diamond microchiplet. SiN waveguides provide optical input (8) and output (8) access to the diamond using 16 total channels. To ease the integration of the PIC module with a commercial optical fiber array, the input and output SiN waveguides were routed to the same side of the PIC and spaced \SI{127}{\micro\meter} apart as shown in Fig.~\ref{fig:gds}(a) of the Appendix. Two SiN waveguide loopback structures that are not routed through the diamond window were added to the same PIC facet to help facilitate alignment with the fiber array and characterize alignment drift during packaging and cryo cooling.

In addition to optical waveguides, metal electrodes were fabricated as part of the PIC module in a \SI{0.75}{\micro\meter}-thick aluminum layer. The role of the electronic capabilities on the PIC is for future microwave control of the qubit and DC strain tuning of the diamond. Ground-signal-ground coplanar microwave waveguides were routed below the diamond window in order to deliver microwave control signals to the SiV${}^-$s. In addition, DC electrodes were patterned in the same aluminum layer and interdigitated with the individual diamond waveguides (see Fig.~\ref{fig:gds}(c)) for future electrostatic actuation-based strain tuning \cite{Meesala2018} of the inhomogeneously-spread color center transition frequencies. For this work, microwave loss was measured but DC bias for strain tuning was not applied.

\begin{figure}[t]
	\includegraphics[width=\linewidth]{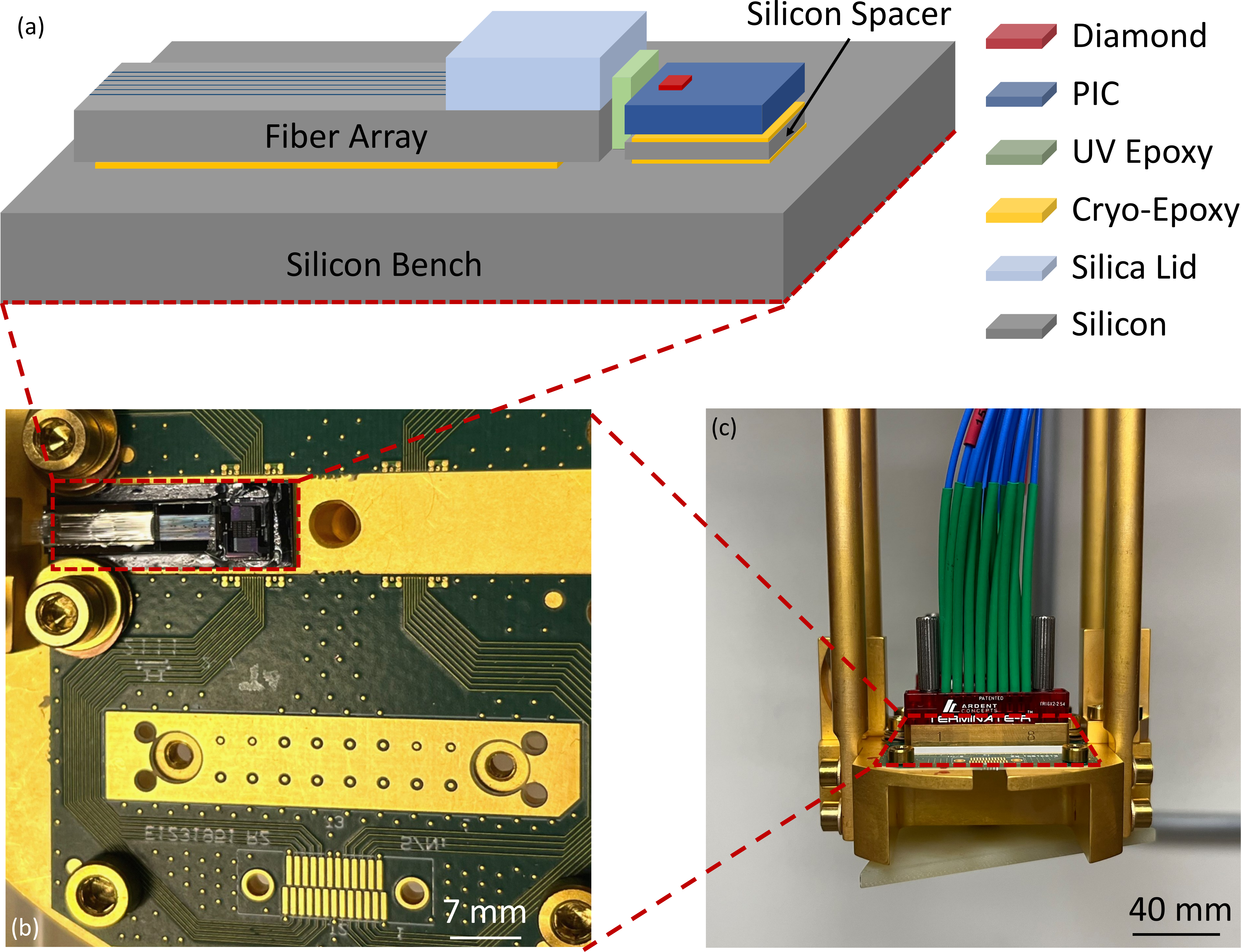}
	\caption{\label{fig:packaging} Packaging details of the quantum-emitter module. (a) The memory module was composed of a silicon bench, a silicon spacer, a silicon fiber array with a pyrex lid, and a silicon-on-insulator PIC. The elements were bonded with cryo-compatible epoxy and UV epoxy was used in the interface. (b) The module was affixed to a gold-plated copper puck, wire bonded to a custom PCB, and (c) mounted in a dilution refrigerator with commercial microwave transmission cables.}
\end{figure}

Attachment of the commercial fiber array to the custom PIC, while maintaining alignment from room temperature down below 100~mK, required the use of thermally compatible materials. We used a common silicon substrate and a silicon spacer---bonded with cryo-compatible epoxy---to match vertical dimensions of the SiN waveguides on the PIC and the optical fibers in the fiber array as shown in Fig.~\ref{fig:packaging}. This majority-silicon design minimized relative motion between the fiber array and the PIC in the module during cooldown. We additionally applied UV epoxy in the interface between the fiber array and the PIC to eliminate any residual alignment drift. While we observed that UV epoxy alone is prone to failure at cryogenic temperatures, the added support from the cryo-compatible epoxy was sufficient to maintain the UV epoxy bond and keep the module aligned. Our packaging procedure compares favorably to other methods (see for example \cite{Mehta2020}). Additional details for the packaging design and procedure are described in Appendix~\ref{app:packaging}.

After the module was fully cured, we placed an eight-channel diamond chiplet onto the PIC (see Fig.~\ref{fig:placement}(b)) using a custom-cut polydimethylsiloxane (PDMS) stamp affixed to a standard glass slide. We used a high magnification zoom lens to image through the stamp to view the diamond chiplet and the PIC simultaneously and positioned the diamond chiplet with a six-axis stage with differential micrometers \cite{Raniwala2023}. Correct diamond alignment with the SiN waveguides can be confirmed at room temperature using photoluminescence (PL, see Sec.~\ref{sec:results} for details). Based on the sub-micrometer width of the SiN waveguides, we estimate a diamond placement accuracy of better than 100~nm, which compares very favorably with the approximately 10~\(\mu\)m placement accuracy inherent to commercially available pick-and-place systems. The module was then attached mechanically to a gold-plated copper puck and wire-bonded to a custom printed circuit board as shown in Fig.~\ref{fig:packaging}. Custom fiber feedthroughs for up to 40 fibers (630HP)---supporting up to 16 diamond channels---through a single KF-40 flange and commercially available microwave and DC feedthroughs enabled operation of the module in our cryogenic system. 

\section{\label{sec:results}Performance Characterization}

\begin{figure}[t]
	\includegraphics[width=\linewidth]{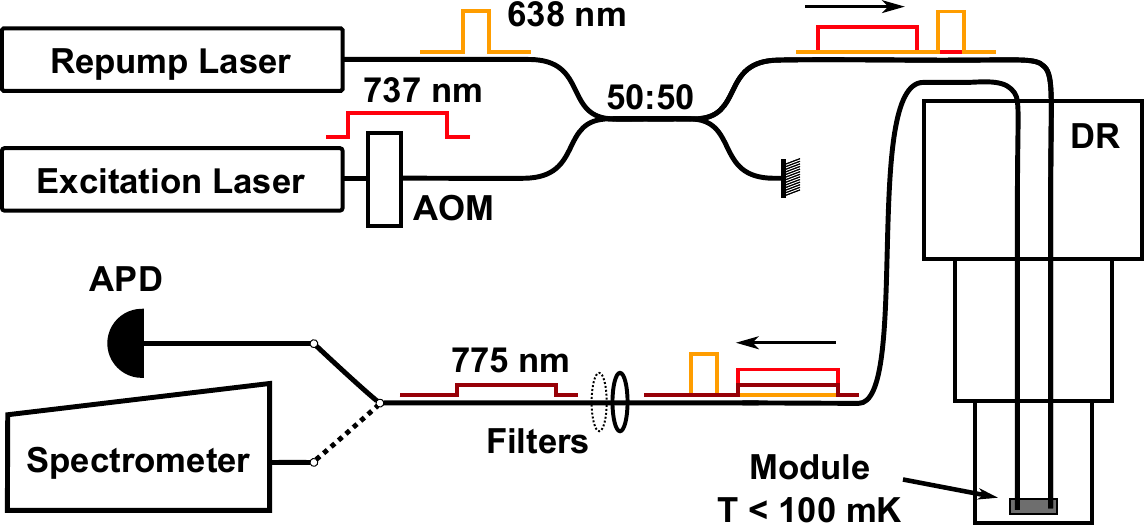}
	\caption{\label{fig:experiment} Experimental setup. The resonant excitation laser light (red, 737~nm) is pulsed using an acousto-optic modulator (AOM), while the off-resonant repump laser (orange, 638~nm) is pulsed using current control. The resonant and off-resonant laser light is combined at a 50:50 fiber splitter/combiner and sent through custom fiber feedthroughs into the dilution refrigerator (DR). The light passes through the emitter module and then exits the DR with co-propagating light from the phonon sideband (brown, 775~nm). The exiting light is filtered using a long pass filter (LPF) that suppresses the the repump laser light and, optionally, a band pass filter (BPF) that passes the phonon side-band. After filtering, the optical signal is sent to a spectrometer to measure photoluminescence or to an avalanche photodiode (APD) to measure photoluminescent excitation.}
\end{figure}

We measured the optical and electrical properties of the quantum-emitter module using the setup shown in Fig.~\ref{fig:experiment}. Light traveled from two laser sources at different wavelengths, combined at a 50:50 fiber splitter/combiner, through an optical fiber and down into a dilution refrigerator (DR). Excitation light was coupled into the module via one channel of a 20-channel fiber array, into the glued PIC and then transitioned from the PIC into the diamond chiplet, driving color centers to emit into the diamond waveguide mode. This fluorescence transitioned back into the PIC waveguide, into the fiber array and out of the DR where it was filtered spectrally and sent to either an avalanche phototodiode (APD) for time-tagged photon detection, or a spectrometer for spectral characterization.

The figures of merit we considered for this architecture include the optical and microwave losses through the packaged system, the detected photon flux from the SiV${}^-$, and the linewidth of the SiV${}^-$ emitters. We characterize the optical insertion loss of the PIC by using a short loopback waveguide structure that does not pass through the diamond window on the PIC. We measured approximately 7~dB round trip insertion loss at 737~nm, which included facet, propagation, and bend losses. Based upon measurements of test devices, we estimate the facet loss to be approximately 3~dB/facet at 737~nm which, crucially, is stable from room temperature down to 100~mK with no active alignment. This compares favorably to the room temperature results of Ref.~\cite{Goede2022} which employs a bi-layer SiN waveguide at 940~nm. The package was cycled from room temperature down below 100~mK multiple times with no appreciable change in transmission. 

The fabricated PIC parameters were swept to explore the optimum geometry for these novel structures and to reveal the most relevant fabrication limitations which will inform future designs. Using cascaded test devices, our best performing window structures had approximately 3.0~dB loss per transition. This loss was primarily due to optical scattering as the waveguide passes directly through window etch which we found to have a rough surface. This fabrication limitation can be overcome by moving to a design with a two-layer SiN stack and adiabatically transitioning between SiN layers within the window region. This would significantly reduce the window transition loss and would also allow for useful capabilities such as polarization control and low-loss optical routing. The best performing SiN-to-diamond transition devices resulted in 6.6 dB loss per transition. Loss in these devices arises from the SiN waveguide layer thickness---which was optimized for edge coupling performance between the PIC and the cleaved fiber array---resulting in imperfect mode matching between the SiN and diamond waveguides.  This limitation can again be addressed by a two-layer SiN stack, where the bottom layer thickness is optimized for cleaved fiber to PIC transmission performance and the top layer thickness is optimized for SiN-to-diamond waveguide transmission. Finally, we estimate an optimum straight waveguide propagation loss of 0.30~dB/cm at 737~nm for 600~nm wide waveguides, which could be improved by increasing the waveguide width from 600~nm up to 700~nm \cite{Sorace2019}. We note that there is increased loss near metals on the PIC, though this can be mitigated by moving the metal layer farther from the optical mode by increasing the oxide depth between the SiN and metal layer.

\begin{figure}[t]
	\includegraphics[width=\linewidth]{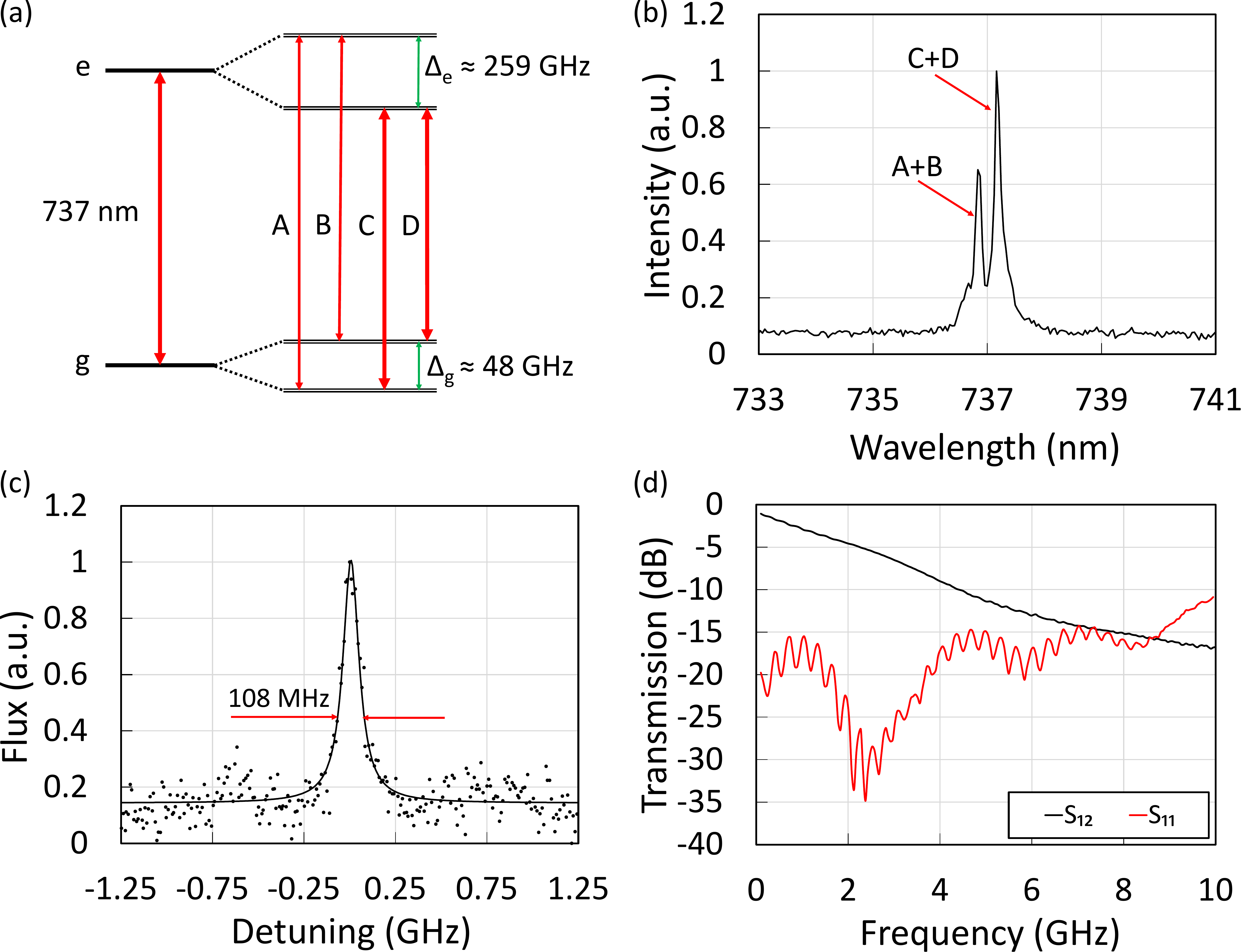}
	\caption{\label{fig:data} Optical and electrical characterization. (a) The energy level diagram of an SiV${}^-$ color center with zero magnetic field. The excited state ($e$) and ground state ($g$) are split by spin-orbit coupling. The resulting manifold contains optical transitions (red lines) and microwave transitions (green lines). (b) A typical PL signal from an ensemble of color centers in a single waveguide using 638~nm excitation light with transitions labeled according to (a). (c) A high resolution PLE scan of one SiV${}^-$ using resonant 737~nm and 638~nm repump laser light and collecting the phonon sideband. Other channels are shown in Fig.~\ref{fig:fullple} in the Appendix. (d) S-parameters of the microwave tranmsission properties of an emitter module.}
\end{figure}

The microwave transmission through a typical module is shown in Fig.~\ref{fig:data}(d).  This data was taken at room temperature, and we have observed minimal changes in microwave transmission characteristics between room temperature and cryogenic operation. The data shows smooth transmission properties out to 10~GHz, indicating the module can deliver distortion free microwave pulses to the  SiV${}^-$ color centers in the module, suitable for enabling quantum memory control operations \cite{Sukachev2017}. The transmission loss is dominated by line-loss from the cables, not by the module itself, which should result in minimal heating of the diamond microchiplet; preliminary results indicate that next generation devices will be able to further reduce microwave loss for improved performance \cite{Christen2022}. 

We demonstrate the functionality of our packaged module by performing optical spectroscopy of six color center emitters in separate diamond waveguides. The SiV energy levels are shown in Fig.~\ref{fig:data}(a), and Fig.~\ref{fig:data}(b) shows a typical photoluminescence (PL) spectrum of an ensemble of color centers in a single waveguide under 638~nm excitation. For these measurements, we used CW light with approximately \SI{5.2}{\micro\watt} of power in the diamond waveguide (calculated based upon known losses) and an integration time of two minutes with a background-subtracted count rate of 2.8 kHz. The output was filtered using a long-pass interference filter at 664~nm. The scan shows two 737~nm transitions due to the excited state splitting of the SiV${}^-$ as shown Fig.~\ref{fig:data}(a). Similar results were found in all six transmitting waveguides. 

To further explore the optical properties of the SiV${}^-$ emitters, we took high-resolution photoluminescent excitation scans of six channels, finding similar results in each as shown in Appendix~\ref{app:PLE}. For these scans, we used the pulse sequence shown in Fig.~\ref{fig:experiment} with optical powers that optimize the signal to noise ratio and linewidth of the emitter. To begin, the repump laser initializes the SiV${}^-$ electronic state with a 500~ns pulse with a peak power of \SI{4.1}{\micro\watt} in the diamond waveguide. After a period of \SI{1.6}{\micro\second}, a \SI{10}{\micro\second} resonant 737~nm pulse with a peak power of \SI{440}{\pico\watt} in the diamond waveguide is applied to cycle the resonant transition. The optical power is calculated based upon known losses. Each full optical pulse cycle is \SI{16.1}{\micro\second} long and includes \SI{10}{\micro\second} of measurement time during the resonant pulse. The output light is filtered to collect the phonon sideband with two 46~nm bandpass filters centered at 775~nm. The resonant laser frequency is swept with emitted phonon sideband photon counts measured by an avalanche photodiode (APD) for many pulse cycles, ranging between $2.50\times 10^5$ and $6.21\times 10^5$ cycles, depending on the channel. A typical scan is shown in Fig.~\ref{fig:data}(c) with an individual SiV${}^-$ resonance including a Lorentzian fit with a full width at half maximum of $108\pm11$~MHz. These results confirm that the packaged module supports individual color center characterization for future use in quantum memory applications.  

\section{\label{sec:conclusion}Discussion and Outlook}

In summary, we have designed, fabricated, packaged, and tested a cryogenically-compatible photonic quantum-emitter module that has the potential to scale to many qubits. In particular, we have resolved two key issues for developing scalable memories for quantum networking---namely, a packaged optical interconnect that is both (a) temperature-stable and (b) compatible with heterogeneous integration techniques, all while maintaining low loss at the PIC-to-fiber interface. 

To put the efficiency of this quantum-emitter module into context, the reported losses associated with successful networked quantum memory demonstrations ranges from tens of dB \cite{Hensen2015} to less than 1~dB \cite{Bhaskar2020}. This work provides an intermediate solution in terms of optical loss at 7 dB, while also demonstrating a scalable architecture with multiple spatially separated emitters. Additionally, we think that the relatively straightforward modification of using a two-layer SiN stack to avoid window transition loss, and the incorporation of lensed fibers to reduce edge coupling loss, will result in a module with approximately 1~dB of loss.

By upgrading our cryogenic setup with a vector magnet to realize a long-lived spin qubit~\cite{Sukachev2017}, and connecting the module to our Boston area quantum network testbed~\cite{Bersin2021, Bunandar2018} we will be able to perform advanced system-level demonstrations across deployed fiber links. Furthermore, while we have designed and demonstrated this module's utility using SiV$^-$ color centers at 737~nm, by modifying its design wavelength our approach can be made compatible with a range of complementary technologies spanning the range of operation of silicon nitride, from the visible to the mid-IR band.  This includes other color centers in diamond~\cite{Trusheim2020}, defects in silicon carbide ~\cite{Anderson2022}, and rare earth ions implanted in various high index materials ~\cite{Stevenson2022}. Additionally, the individual components and techniques developed here have use in enabling efficient novel emitter characterization for quantum information processing use~\cite{Christen2022}, and will likely be of significant benefit to interconnects between disparate technologies for quantum~\cite{Awschalom2021,Najafi2015} and classical applications~\cite{Stern2013, Hermans2021,Chatzipetrou2021}.

\begin{acknowledgments}
Distribution Statement A. Approved for public release. Distribution is unlimited. This material is based upon work supported by the National Reconnaissance Office and the Under Secretary of Defense for Research and Engineering under Air Force Contract No. FA8702-15-D-0001. Any opinions, findings, conclusions or recommendations expressed in this material are those of the authors and do not necessarily reflect the views of the National Reconnaissance Office or the Under Secretary of Defense for Research and Engineering. \copyright~2023 Massachusetts Institute of Technology. Delivered to the U.S. Government with Unlimited Rights, as defined in DFARS Part 252.227-7013 or 7014 (Feb 2014). Notwithstanding any copyright notice, U.S. Government rights in this work are defined by DFARS 252.227-7013 or DFARS 252.227-7014 as detailed above. Use of this work other than as specifically authorized by the U.S. Government may violate any copyrights that exist in this work.
 
This material is based upon work supported by the Center for Quantum Networks under NSF grant 1941583. M.S. acknowledges support from the NASA Space Technology Graduate Research Fellowship Program. Work performed on MIT campus was supported in part by the Air Force Office of Scientific Research under award number FA9550-20-1-0105, supervised by Dr. Gernot Pomrenke.
\end{acknowledgments}

\appendix

\section{DIAMOND FABRICATION}
\label{app:diamondfab}

The diamond micro chiplet was fabricated using a similar process to the one shown in Refs.~\cite{Mouradian2017, Wan2018, Wan2020}. The fabrication process began with an electronic grade single-crystal diamond plate from Element 6. We performed plasma etching in Ar/Cl${}_2$ followed by O${}_2$ etching to remove \SI{7}{\micro\meter} of the diamond surface in order to relieve strain in the sample. This was followed by ion implantation of ${}^{29}$Si by  Innovation Inc.~with an effective areal dose of $10^{11}$~ion/cm${}^{2}$ at 175~keV for a mean depth of 115~nm. After implantation, the sample was annealed at \SI{1200}{\degreeCelsius} in an ultra-high vacuum furnace at $10^{-7}$ mbar, followed by cleaning in a boiling mixture of nitric acid, sulfuric acid, and perchloric acid (1:1:1) to remove any graphite formed during the anneal. 

Once this implantation process was complete, diamond microchiplets were fabricated following the methods described in Ref.~\cite{Mouradian2017}. The chiplet is shown in Fig.~\ref{fig:gds}(c) and (d). The diamond waveguides taper up linearly from approximately 50~nm to 320~nm over \SI{3.0}{\micro\meter}. The eight diamond waveguides have a rectangular shape with a depth of 200~nm and are spaced be \SI{3.0}{\micro\meter}. The resulting suspended microchiplet was then removed with a tungsten probe and placed (smooth-side-up) onto a custom polydimethylsiloxane (PDMS) stamp to transfer (smooth-side-down) to the SiN PIC, as described in Sec.~\ref{sec:package}.  

\section{PIC FABRICATION}
\label{app:PICfab}

PICs were used to photonically access diamond color centers by routing optical signals at 628~nm and 737~nm from optical fiber, through SiN waveguides, and into a diamond micro-chiplet. These PICs were fabricated at wafer scale using MIT Lincoln Laboratory’s 90~nm node 200~mm wafer foundry, using a SiN-on-SiO${}_2$ plasma enhanced chemical vapor deposition (PECVD) process \cite{Sorace2019}.

For our devices, an aluminum layer (\SI{0.75}{\micro\meter} thickness) was deposited on SiO${}_2$ for future microwave and DC control of SiV${}^-$ color centers, followed by \SI{1.5}{\micro\meter} of SiO${}_2$ PECVD cladding. The oxide is then planarized via chemical-mechanical polishing to a total depth of \SI{0.6}{\micro\meter}. A 100-nm-thick SiN layer was deposited and etched to form the optical waveguides. Due to the lack of a SiN-SiO${}_2$ selective etch \cite{Williams2003}, we employed a titanium-aluminum (Ti-Al) etch-stop layer using a liftoff process. The etch-stop layer was deposited on top of selected SiN waveguides to aid in the removal of oxide cladding in regions where we employ diamond integration. This was followed by depositing \SI{5}{\micro\meter} of SiO${}_2$ and a standard masked oxide etch to open a window down to the etch-stop layer, followed by a wet etch that removed the Ti-Al etch stop layer, leaving the SiN layer underneath. Vias were patterned and etched to access the metal layer for wire bonding, and a \SI{150}{\micro\meter}-deep and \SI{100}{\micro\meter}-wide trench was used to define the dicing trenches on four sides of the entire $4\times5$~mm PIC. An additional \SI{500}{\micro\meter}-deep etch was done on the optical input facets of the PIC, resulting in a total etch depth at the facet of \SI{650}{\micro\meter}. This enabled the fiber array to come flush against the PIC facet during the bonding process. The resulting PICs had both clad and unclad SiN waveguides, high quality edge facets and embedded microwave and DC electrodes. The fabrication steps relevant for opening the diamond window are shown in Fig.~\ref{fig:fab}.

Photonic structures on the PIC module included linear inverted tapers, bends, and evanescent couplers to get the optical mode in and out of the diamond. This latter structure is composed of two parts: (a) a ``window transition region'' composed of a taper from the clad SiN waveguide to the unclad SiN waveguide in the open window region of the device; and (b) a ``SiN-to-diamond transition'' region composed of a linear taper from an unclad SiN waveguide into a diamond waveguide. The details are shown in Fig.~\ref{fig:gds}(d-e). The window transition is accomplished by tapering the SiN waveguide from a 600~nm width in the clad region---section (i) in Fig.~\ref{fig:gds}(e)---to a 1000~nm width in the unclad region---section (ii). During this taper, the window tapers open gradually over a distance of \SI{5}{\micro\meter}---sections (ii) to (iii). The SiN-to-diamond transition is accomplished in the unclad window region, where the SiN waveguide is tapered down while the diamond waveguide is tapered up towards the desired width of 320~nm---section (iv). The optical mode is thus evanescently coupled from the 1000~nm-wide SiN waveguide to the diamond waveguide---section (v)---where it can interact with SiV${}^-$ color centers. The losses of the window transition were measured to be 3.3~dB/transition using a variation of the cut-back method, where several back-to-back window transitions were nested together and the insertion loss over multiple window transitions was measured. The losses of the SiN-to-diamond transition were estimated to be 6.6~dB/transition by measuring total PIC insertion losses with diamond attached and back-calculating the diamond transition losses from known insertion losses of all other components. A standard parameter sweep on separate test structures allowed identification of designs with lower losses, to be used as part of the full module design in future tape-outs. We note that although Fig.~\ref{fig:gds}(d) shows the layout of an ideal diamond placement onto the etched window, in practice small angular displacements of the diamond microchiplet with respect to SiN waveguides resulting from manual diamond positioning could result in higher insertion losses.

Reasonable changes to the design and fabrication of these devices can offer significant improvements to the loss in the transition to the unclad SiN and from the unclad SiN waveguide to the diamond waveguide. These improvements can be accomplished by employing two layers of SiN and optimizing the layer thicknesses and taper parameters of the SiN and diamond waveguides. 

\section{PACKAGING}
\label{app:packaging}

As shown in Fig.~\ref{fig:packaging}, a \SI{725}{\micro\meter} thick silicon bench formed a common substrate for the PIC and fiber array; however, a \SI{275}{\micro\meter} thick silicon spacer between the silicon bench and the PIC was needed to allow clearance for the fiber core to align with the facet of the PIC. We designed a custom jig to ensure precise alignment of the bench, spacer and PIC. Using this jig, along with custom cut spacers made from plastic shim stock, we were able to maintain a bond line of cryo-compatible epoxy (MasterBond EP21TCHT-1) measuring approximately \SI{76}{\micro\meter} for each layer. Pressure was applied with a Teflon-tipped screw from above for approximately two hours to ensure that the PIC, spacer and bench were parallel. 

\begin{figure}[t]
	\includegraphics[width=\linewidth]{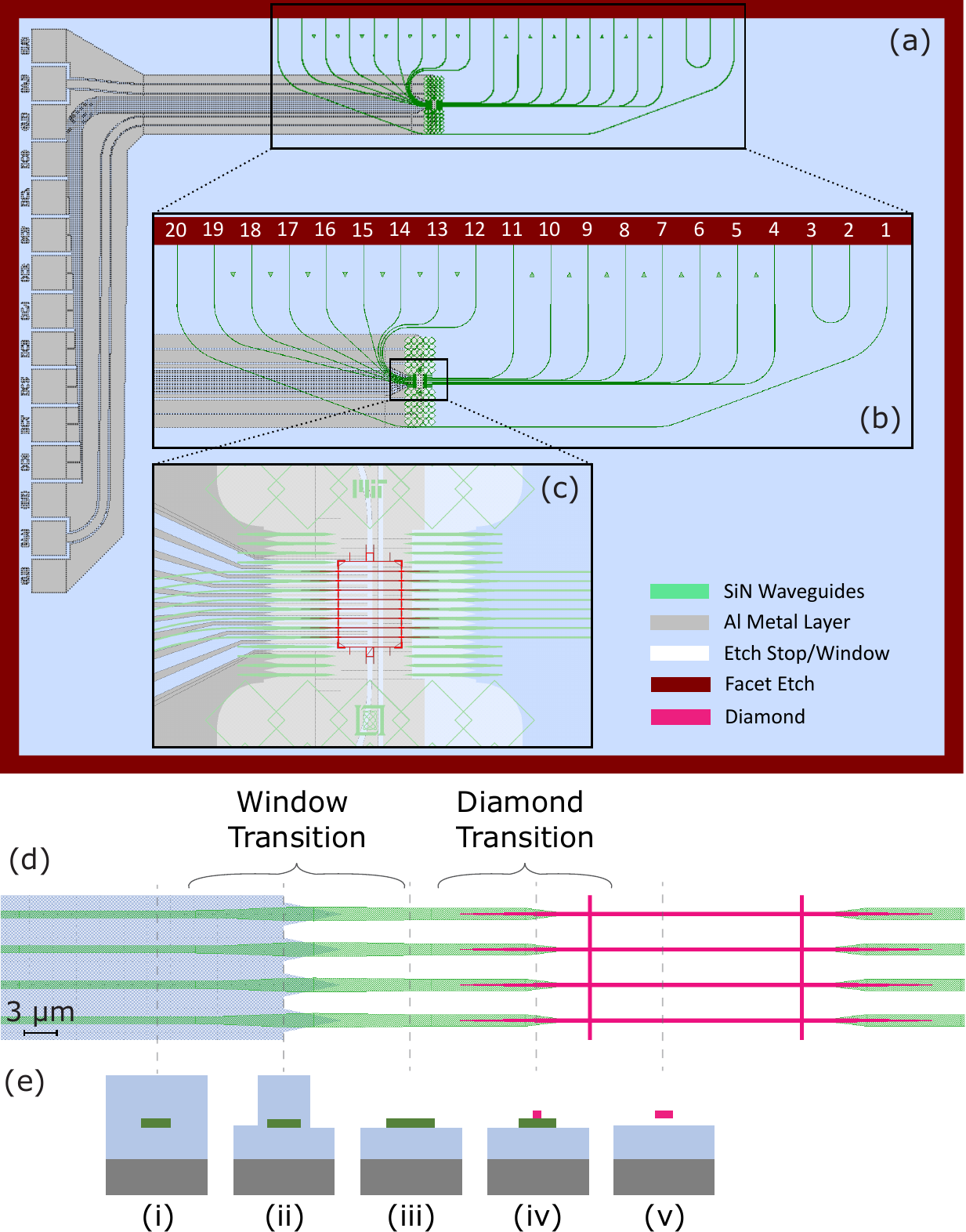}
	\caption{\label{fig:gds} PIC and diamond layout. (a) A layout of the full PIC with relevant structures, including two magnified insets. (b) The facets are shown with 10 optical input and 10 optical output channels. Channels 1, 2, 3 and 20 are used for optical alignment of the 20-channel fiber array, while channels 4-19 are used for diamond access. (c) The diamond integration region with diamond shown. (d) A magnified view of the window and diamond transition regions. (e) A cross section view of (d), with regions (i)-(v) detailed in the text.}
\end{figure}

\begin{figure}[ht]
	\includegraphics[width=\linewidth]{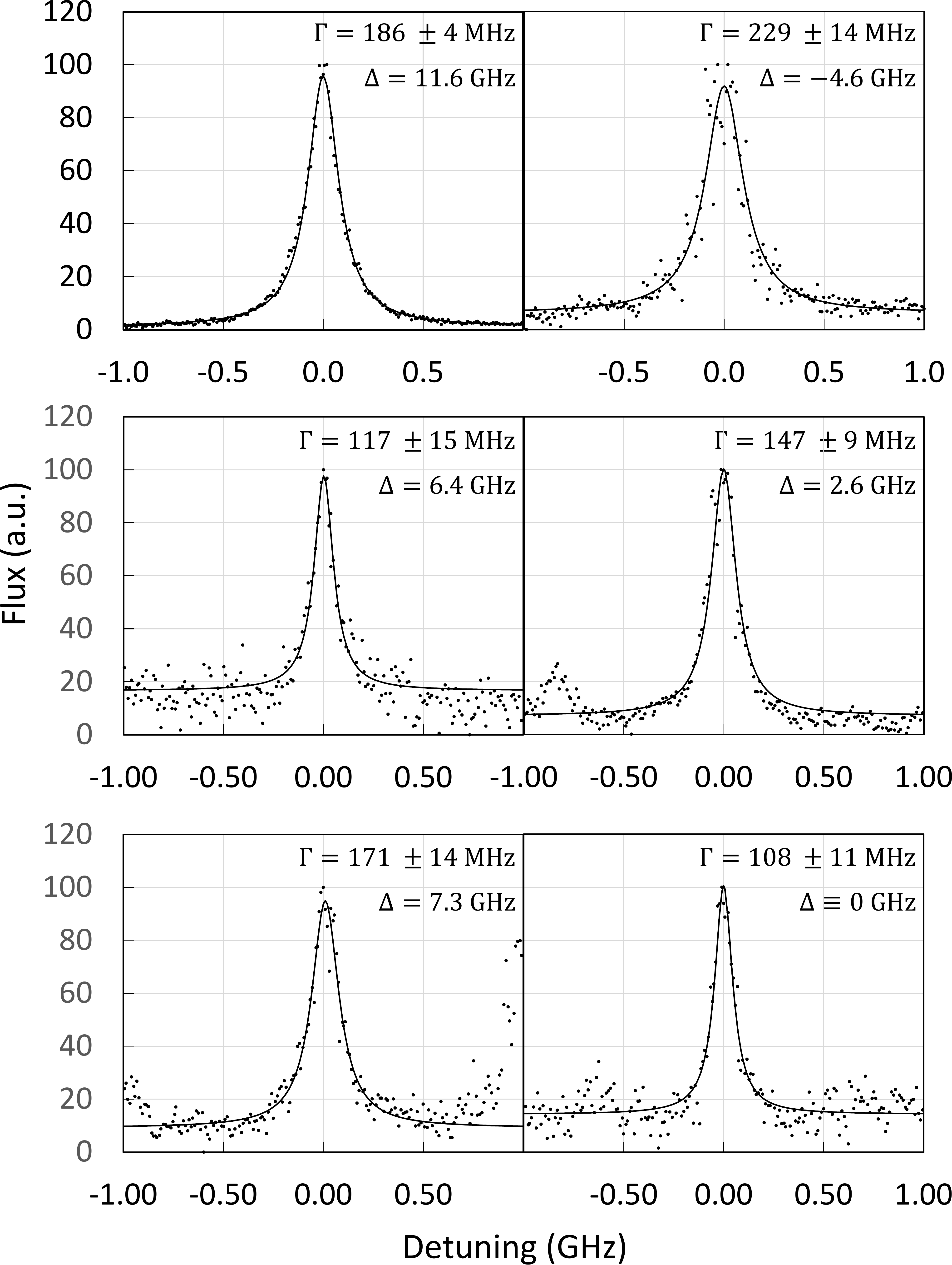}
	\caption{\label{fig:fullple} Photoluminescent excitation (PLE) results. Six diamond channels were measured via the methods explained in the text. Lorentzian full width at half maximum $\Gamma$ and the optical shift relative to the channel in Fig.~\ref{fig:data}(c) is shown.}
\end{figure}

Before the epoxy was fully cured, we moved the package to a bonding platform consisting of a three-axis translation stage and a custom designed Delrin fixture to temporarily mount the PIC stack-up. A commercial 20-channel silicon fiber array composed of 630HP single mode fiber was mounted to a six-axis translation stage with differential micrometers and a custom fabricated rigid fixture. This fixture provided strain relief to the fibers and was able to release the fiber array after curing without disturbing the alignment. A high magnification optical microscope with viewing camera was used to aid in bonding. Using a 730~nm diode laser with a polarization controller, we optimized alignment through loopback alignment structures on the PIC (see Fig.~\ref{fig:gds}(b)), ensuring accurate alignment across the full length of the facet. Once aligned, the fiber array was raised out of the way and epoxy was applied to two regions: a) the cryo-compatible epoxy was applied to the silicon bench beneath the fiber array; b) UV epoxy (Norland 63) was applied to the facet of the fiber array. The fiber array was then lowered back into position, forming a bond both below the array and between the array and the PIC. The alignment was optimized and then UV light was applied to the interface for approximately 60 minutes to ensure a complete cure. Compared to other methods \cite{Mehta2020}, this bonding procedure is completed without the need for a temperature controlled alignment stage. We allow the module to cure in the alignment setup for 24 hours before transferring to a dry box to complete the room temperature cure of the cryo-compatible epoxy over an additional 48 hours. 

\section{OPTICAL CHARACTERIZATION}
\label{app:PLE}
We measured PLE signals on six diamond channels as discussed in Sec.~\ref{sec:results}. Data was taken using the methods discussed there, with integration times varied from 2.50~s up to 6.20~s ($2.50\times 10^5$--$6.21\times 10^5$ cycles) to ensure adequate signal-to-noise based upon the brightness of the individual emitter. The results are shown in Fig.~\ref{fig:fullple} with an arbitrary vertical axis. The Lorentzian full width at half maximum $\Gamma$ as well as the shift in center frequency $\Delta$ relative to the Fig.~\ref{fig:data}(c) is shown. All emitters are within a 12 GHz range, and the linewidths vary from 108 MHz up to 229 MHz.

\bibliography{MemoryModule2023}

\end{document}